\documentclass{aa}
\usepackage{txfonts}
\usepackage{graphicx}

\begin{document}

\title{Different Sodium enhancements among multiple populations of
Milky Way globular clusters}

\author{Andr\'es E. Piatti\inst{1,2}\thanks{\email{andres.piatti@unc.edu.ar}}}

\institute{Instituto Interdisciplinario de Ciencias B\'asicas (ICB), CONICET-UNCUYO, Padre J. Contreras 1300, M5502JMA, Mendoza, Argentina;
\and Consejo Nacional de Investigaciones Cient\'{\i}ficas y T\'ecnicas (CONICET), Godoy Cruz 2290, C1425FQB,  Buenos Aires, Argentina\\
}

\date{Received / Accepted}

\abstract{We searched for trails to understand the different Na abundances measured
in first and second generation stars of ancient Milky Way globular clusters. For that
purpose, we gathered from the recent literature the aforementioned Na abundances, 
orbital parameters, structural and internal dynamical properties and ages in an homogeneous
scale of 28 globular clusters. We found that the intra-cluster Na enrichment, 
measured by the difference of Na abundances between first and second generation stars,
exhibits a trend as a function of the Na abundances of first generation stars, in the sense
that the more Na-poor the first generation stars, the larger the Na enrichment. By
using the inclinations of the globular clusters' orbits, the analyzed Na enrichments also 
hinted at a boundary at $\sim$ 0.3 dex to differentiate globular clusters with an accreted or in situ
origin, the accreted globular clusters having larger Na enrichments. Because 
relatively larger intra-cluster Na enhancements are seen in accreted globular clusters,
and small Na enhancements are observed in globular clusters formed in situ (although not
exclusively), we speculate with the possibility that the amplitude of the Na enrichment could be linked with  the building block paradigm. Globular clusters at the time of formation of first 
and second generation stars would seem to keep memory of  this hierarchical galaxy formation
process.}
 

\titlerunning{Multiple populations of globular clusters}
\authorrunning{Andr\' es E. Piatti }

\maketitle

\markboth{Andr\' es E. Piatti: Multiple populations of globular clusters}{}

\section{Introduction} 

Multiple populations is a phenomenon commonly observed in Milky Way globular clusters,
which exhibit stellar populations with distinctive chemical  abundance patterns \citep{grattonetal2004,bl2018}. Among the chemical elements that witness such a wide range of
values, Na has become the flagship. This is because it has been measured in every stellar
aggregate harboring multiple populations, so that intrinsic Na spreads have been used as an observational evidence. The mechanism that triggers the enhancement of Na, and light elements in general,
is still under debate; a summary of them can be found in \citet{wangetal2020}. The most
frequently discussed scenarios suggest the existence of polluters inside the clusters, responsible 
for the chemical enhancement, namely: intermediate-mass asymptotic giant branch stars, fast
rotating massive main sequence stars, interacting massive binaries, and super massive
main sequence stars. These polluters would enrich the intra-cluster medium in a remarkably
short space of time - from nearly zero to the order of tens up to a couple of hundred Myrs 
- compared with the globular clusters' ages \citep{oliveiraetal2020,saracinoetal2020,cs2020}. 

Recently, \citet{marinoetal2019} compiled Na abundances for first and second generation
stars in 28 Milky Way globular clusters. From the position of these stars in the chromosome 
map \citep{miloneetal2017},
they identified two types of globular clusters, called Type I and Type II globular clusters. More than 80\% of the
globular clusters studied by them turned out to be of Type I, whose chromosome maps look much
simpler than those of Type II. With the aim of analyzing the global properties of multiple populations, they built a
universal chromosome map, which revealed a tight connection with Na abundances. First
generation stars are usually associated to those formed at the time of the globular cluster
formation, so that their
chemical compositions are subject to the metallicity content of the cloud out of which the
globular cluster was formed. Since some globular clusters were formed in dwarf galaxies
later accreted by the Milky Way, while others formed in situ \citep{kruijssenetal2019}, the 
Na abundances of  their first generation stars should follow the  galactic chemical enrichment 
at the time of galaxy formation. Indeed, Na abundances of first generation stars in
the globular clusters studied by \citet{marinoetal2019} span a range of values. 

As far as we are aware, there is no explanation for the level of Na abundances of second
generation stars, which also shows a range of values among globular clusters. Precisely, in this 
work we address this issue, with the aim of providing some clues on the origin of second
generation stars. In Section 2, we describe the data we gathered in order to carry out
our analysis, while in Section 3 we discuss our findings in the light of a cosmological 
context.

\section{The data}

We make use of four different pieces of information: the Na abundances compiled by
\citet{marinoetal2019}; dynamical properties, such as the semi-major axis, the eccentricity, 
the inclination of the globular clusters' orbits, and their space velocity components ($V_r$,
$V_\theta$, $V_\phi$), taken from \citet{piatti2019}; structural and internal dynamics 
evolutionary properties (e.g., half-mass relaxation times, the ratio of the cluster
mass lost by tidal disruption to the total cluster mass, the Jacobi radius) computed by
\citet{baumgardtetal2019} and \citet{piattietal2019b}; and the globular clusters' ages
homogeneously obtained by \citet{valcinetal2020} using the same method and put them
in the same age scale. We note that the inclination of the globular clusters' orbits ($i$)
ranges from 0$\degr$ for fully prograde in-plane orbits to 90$\degr$ for polar orbits
to 180$\degr$ for in-plane retrograde orbits.

Following  \citet{fb2010}'s precepts, we consider retrograde motions being the signature of 
globular clusters that have been accreted in the opposite rotational sense to the main bulk of Milky 
Way's rotation. We note, however, that accreted globular clusters can also have prograde 
orbits. For this reason, \citet{fb2010} also investigated the age-metallicity relationship 
as a diagnostic tool to disentangle accreted and formed  in situ globular clusters. We adopted 
here the results obtained by \citet{piatti2019} from the analysis of the distribution  of $i$ values 
of 156 Milky Way globular clusters, who found a similar number of accreted globular clusters 
with prograde and retrograde orbits. In the subsequent analysis, we bear in mind that among
the  28 globular clusters analyzed here, there could be a similar number of
accreted globular clusters with prograde orbits as that with retrograde ones. 

As far as the completeness of the globular cluster sample is concerned, \citet{marinoetal2019}
pointed out that they described the universal properties of globular clusters in the chromosome
map, so that any globular cluster can be found with Na abundances for first and second
generation stars within the quoted ranges. In this sense, the analyzed Na abundances
are representative of those for the entire Milky Way globular cluster population. In what follows,
we will call [Na/Fe] for first and second generation stars simply by1G and 2G, respectively.

\begin{figure}
\includegraphics[width=\columnwidth]{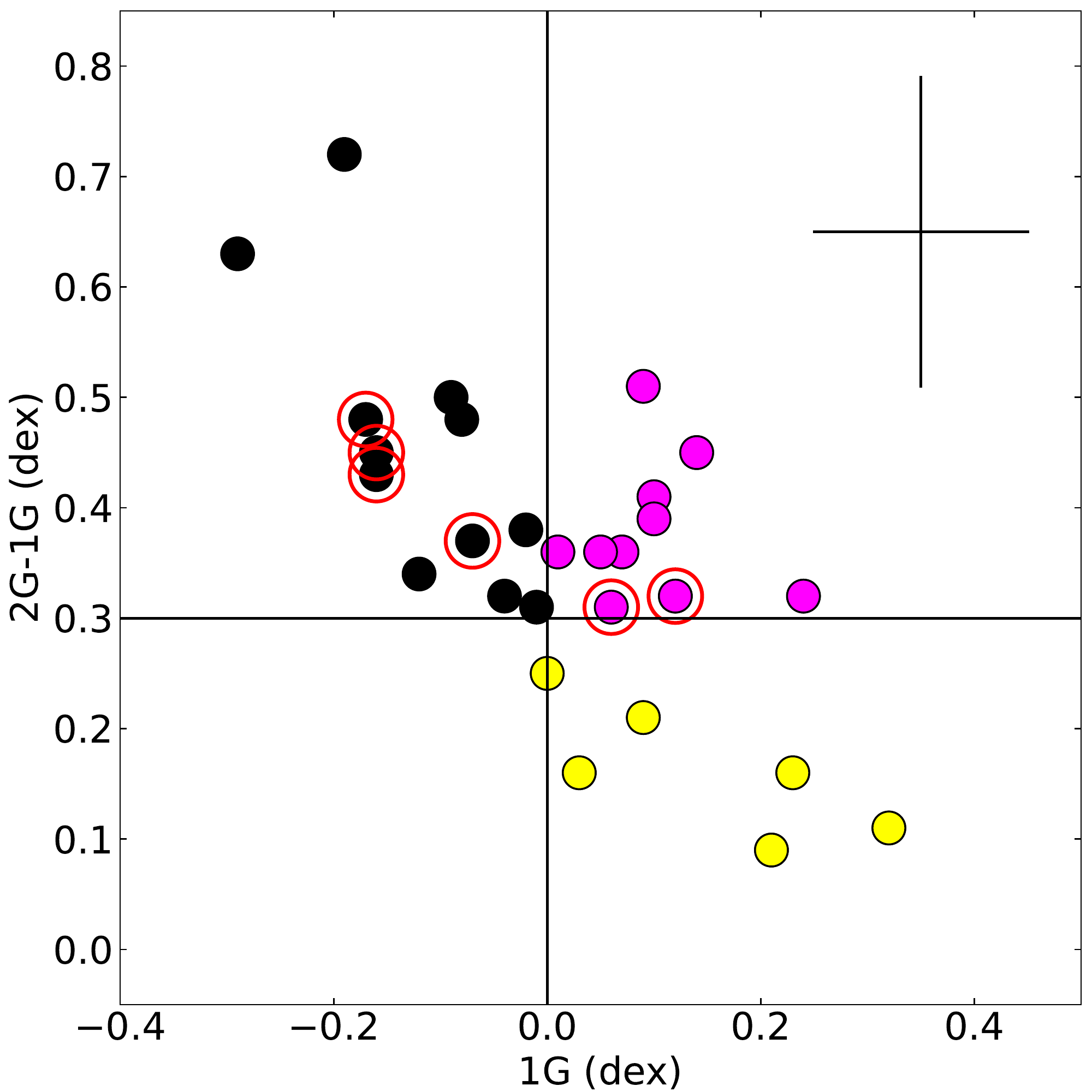}
\caption{The difference of Na abundances between first and second generation stars versus
that for the first generation stars. Typical error bars are included. Black, magenta, and yellow
circles represent globular clusters located in three different quadrants, defined by 2G-1G = 0.3
dex and 1G = 0.0 dex. Large open red circles represent Type II globular clusters.}
\label{fig:fig1}
\end{figure}

\section{Analysis and discussions}

As shown by \citet{marinoetal2019}, there are Na abundance spreads among Milky
Way globular clusters. Figure~\ref{fig:fig1} shows that the intra-cluster enhancement in the
Na abundances (2G-1G) is not random, but follows a trend with 1G, in the sense that
the more Na-poor the first generation stars, the higher the Na enhancement. In the figure, we 
distinguish four quadrants defined by the horizontal line at 2G-1G = 0.3 dex and the
vertical line at 1G = 0.0 dex.   We note that no globular cluster occupies the quadrant 
delimited by 2G-1G < 0.3 dex and 1G < 0.0 dex. Pal\,6 has a Na abundance of 
-0.46$\pm$0.02 dex and is the first convincing example of a single-population globular cluster,
although its present mass (log($M$ /$M_\odot$)=4.83) is much higher than the lower mass
limit of globular clusters with multiple populations \citep{villanovaetal2013,cs2020}. By
adopting 2G-1G=0.0 dex for Pal\,6, we find that it falls outside the range of 1G values
for globular clusters with multiple populations. The different levels of 1G -- analogs to field stars --
shows that globular clusters were formed in environments with different primordial
Na abundances.

\begin{figure}
\includegraphics[width=\columnwidth]{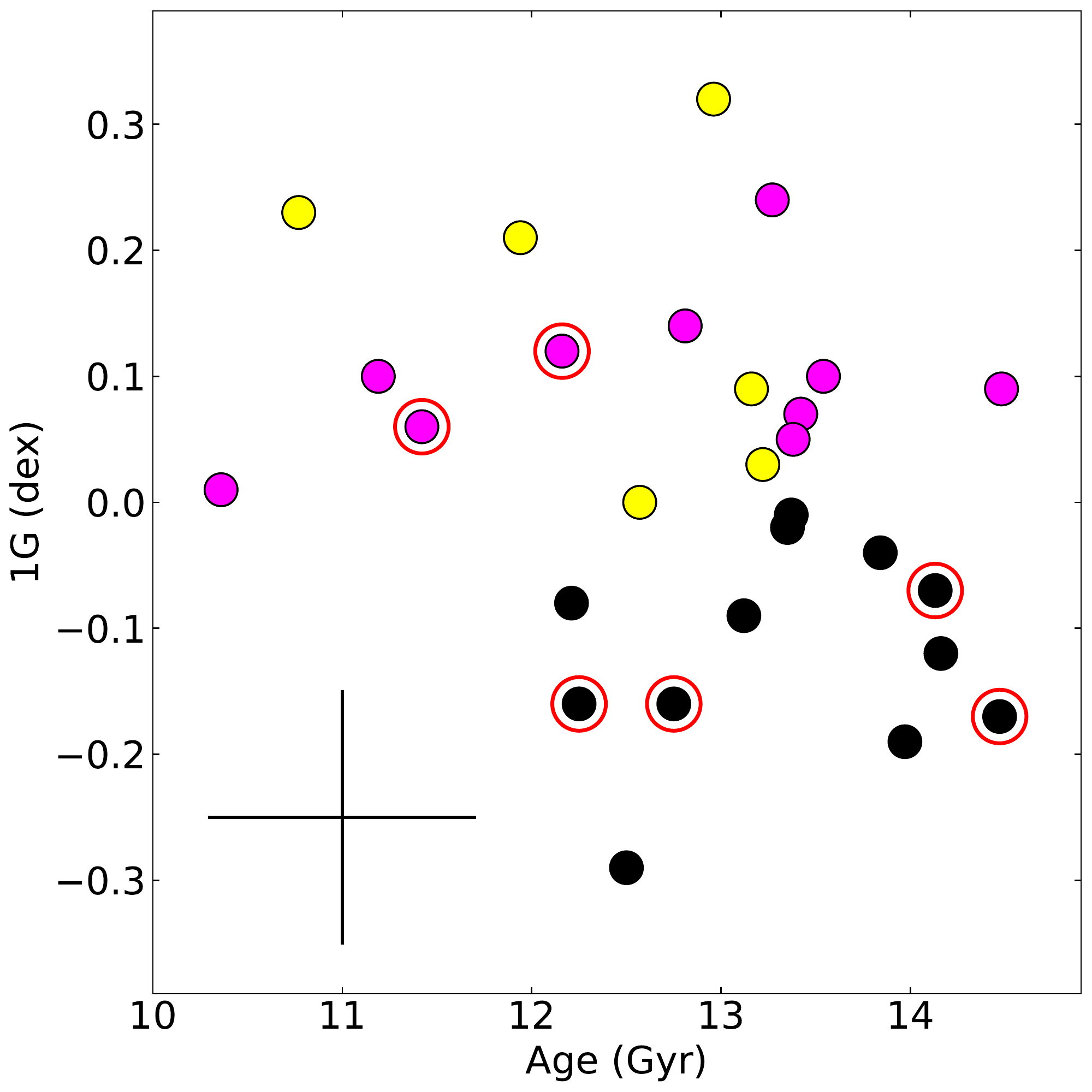}
\caption{Na abundance of first generation stars versus the globular clusters' ages. Typical error 
bars are included. Symbols are as in Figure~\ref{fig:fig1}.}
\label{fig:fig2}
\end{figure}

It has been shown that the most Na-poor limit in dwarf galaxies is lower than the Na abundance 
of Milky Way field stars, with some exceptions \citep{coluccietal2012,ishigakietal2014,battagliaetal2017,villanovaetal2019,salgadoetal2019,matsunoetal2019,aguadoetal2020}. Therefore, first generation stars 
of  globular clusters formed in accreted dwarf galaxies should mostly have Na abundances lower
than their counterparts of globular clusters formed in situ, which explains the range of 1G 
values seen in Figure~\ref{fig:fig1}. The Na abundance of field stars formed  in situ is nearly 
0.0 dex \citep[see, e.g.,][]{hilletal2019}, so that we can assume 1G = 0.0 dex as a first guess for 
a representative boundary to differentiate accreted from formed in situ globular
clusters. We note, however, that there could be accreted globular clusters with Na abundances
of first generation stars similar to that of globular clusters formed in situ. With this
distinction, black circles in Figure~\ref{fig:fig1} correspond to globular clusters with an accreted
origin, as well as some of the magenta ones (see discussion below).

According to the building block paradigm \citep{wr1978,fontetal2011}, dwarf galaxies formed in an earlier Universe are
expected to be older and more chemically deficient than galaxies formed from the assembly of
those primordial dwarfs. Figure~\ref{fig:fig2} shows that the 1G values, which refer to the
most  Na-poor values of the galaxies where the globular cluster were formed, hints at an
age-Na abundance relationship in agreement with the mentioned galaxy formation scenario. 
As can be
seen, globular clusters younger than 12 Gyr are more Na-rich than 0.0 dex, while the
most Na-poor globular clusters (1G < 0.0 dex) are among the oldest ones. Nevertheless,
there are old globular clusters with Na-rich values (1G > 0.0 dex), which somehow reveals
 that the Na enrichment was more intense during the first $\sim$ 2 Gyr.
We point out that, because of the relative short space of
time between the formation of first and second generation stars, second generation stars in 
accreted globular clusters have been formed beforehand the globular clusters were accreted 
to the Milly Way. This means that the difference 2G-1G is a measure of the intra-cluster Na
enhancement at the time of the globular cluster formation.

\begin{figure}
\includegraphics[width=\columnwidth]{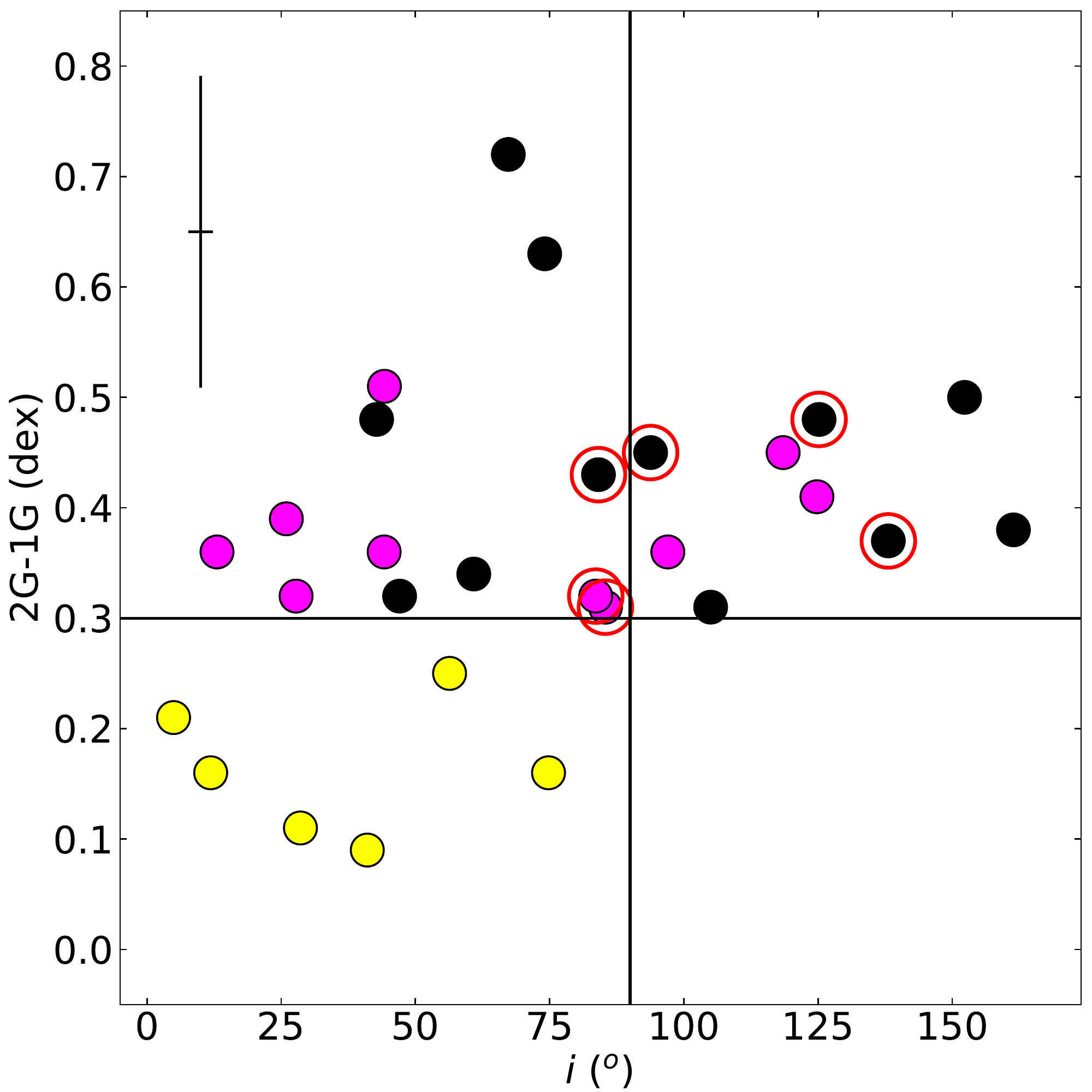}
\caption{The difference of Na abundances between first and second generation stars versus
the inclination of the globular clusters' orbits. Typical error bars are included. Symbols are as
in Figure~\ref{fig:fig1}.}
\label{fig:fig3}
\end{figure}

\begin{figure}
\includegraphics[width=\columnwidth]{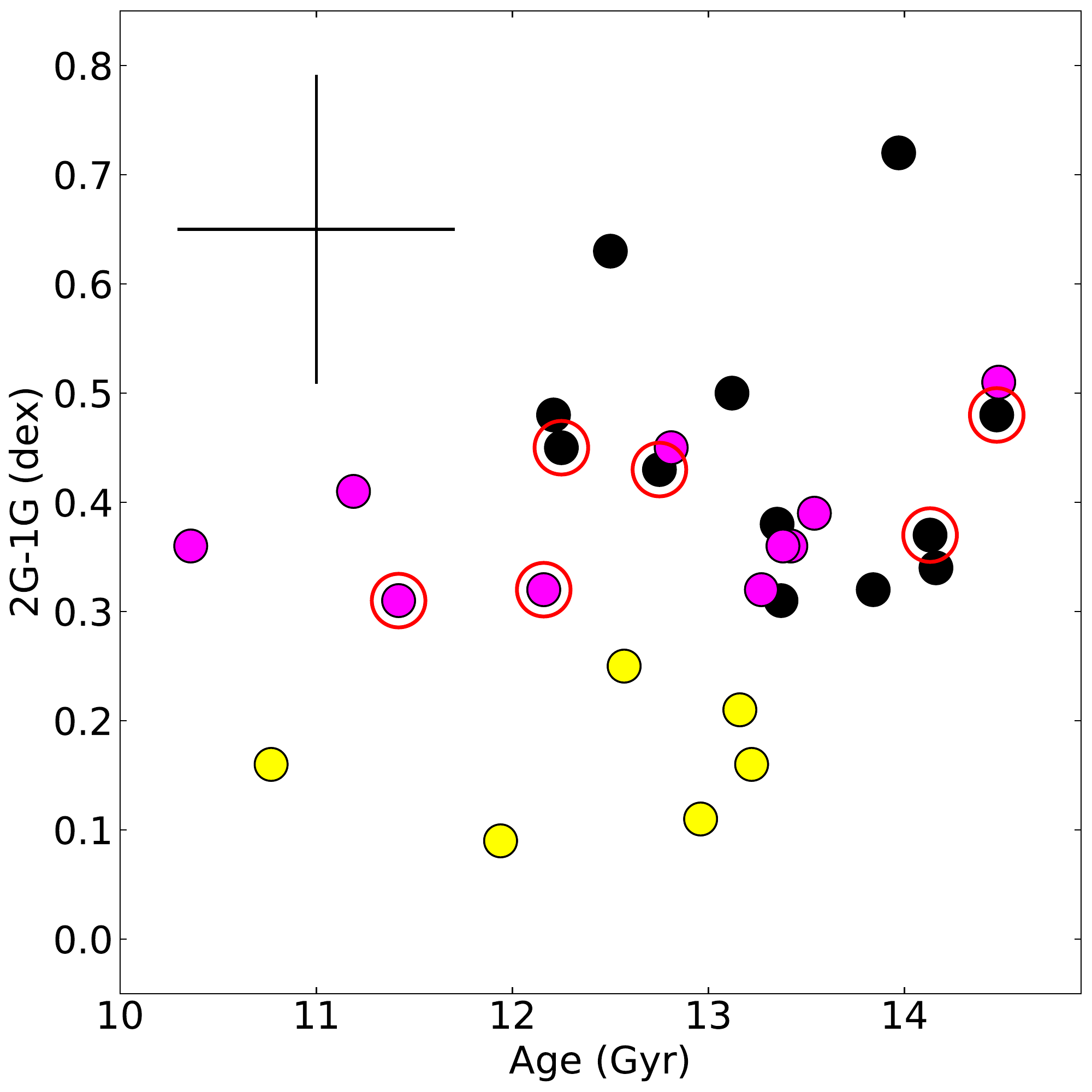}
\caption{The difference of Na abundances between first and second generation stars versus
the globular clusters' ages. Typical error bars are included. Symbols are as in Figure~\ref{fig:fig1}.}
\label{fig:fig4}
\end{figure}

We played with the different globular cluster parameters mentioned in Section 2 and found that
the inclination of the globular clusters' orbits can help in recognizing accreted from formed
in situ globular clusters.  The remaining astrophysical parameters 
do not show any clear correlation with 1G nor with 2G-1G (see Appendix). Figure~\ref{fig:fig3}
shows that globular clusters with retrograde
orbits ($i$ > 90$\degr$), and hence with an accreted origin, have 2G-1G > 0.3 dex. These
globular clusters are Na-poor (1G < 0.0 dex, black circles in Figure~\ref{fig:fig1}) and 
older than 12 Gyr (see Figure~\ref{fig:fig2}), or old and Na-rich ones. In either case, their
orbital inclinations, ages and Na abundances agree well with having formed in accreted dwarf
galaxies. Therefore, we adopt 2G-1G = 0.3 dex as a boundary to differentiate globular
clusters with accreted or in situ origins. The top-right quadrant of Figure~\ref{fig:fig3}
contains accreted globular clusters, while the bottom-left one, only globular clusters
formed  in situ. In the top-left quadrant are located globular clusters with either
accreted or  in situ origin. According to \citet{piatti2019}, half of the accreted globular
clusters could have prograde orbital motions  ($i$ $<$ 90$\degr$). In this sense, the older 
and more Na-poor globular clusters (black circles) in the top-left quadrant would correspond
to accreted globular clusters (see also Figure~\ref{fig:fig1}), 
as well as some of the magenta ones. We note that it is not possible to assess on the
globular cluster origin using Figure~\ref{fig:fig3} for those with 1G > 0.0 dex,
2G-1G > 0.3 dex, and ages older than 12 Gyr. Figure~\ref{fig:fig1} can also be use as a 
complementary diagnostic diagram: top-left and bottom-right quadrants correspond to accreted 
and formed in situ globular clusters, respectively, while the top-right panel can contain a 
mixture of them. The trend shown in the figure can now be interpreted in terms of the 
frequently referred cosmological hierarchy of galaxy formation. The oldest globular clusters 
formed in primordial dwarf galaxies with very deficient Na abundances, while those formed
in the Milky Way took more Na-rich values. 

Globular clusters represented with black and yellow circles in Figure~\ref{fig:fig1}  not only
have different 1G values but also Na enhancements (2G-1G).
A possible interpretation for such a difference between these two groups of globular clusters
can be drawn from Figure~\ref{fig:fig4}, where the Na enhancements are plotted as a function
of the globular clusters' ages. Figure~\ref{fig:fig4} reveals some broad correlation, in the sense
that the older globular clusters, the higher the intra-cluster Na enhancement. As discussed
above,  most of the black and some magenta circles could represent globular clusters formed
in dwarf galaxies that later were involved in the assembly of the Milky Way. Therefore,
 some globular clusters formed inside these first galaxies  could have experienced more vigorous enhancement
processes during their formation that resulted in a wider range of Na abundances, as compared
with  most of the globular clusters formed in situ.
In other words, the Na enrichment inside globular clusters
would seem to have been more efficient in the early Universe than at the time of formation
of globular clusters in the Milky Way. This picture leads us to speculate about some kind of loss 
of strength or deceleration of the Na enrichment process inside the globular clusters. If
second generation of stars were the result of the interaction of binary stars, or from the ejecta
of asymptotic giant branch stars, or fast rotating main sequence stars, or stellar mergers,
among others \citep{wangetal2020}, the above findings would imply that those processes
would have lost  effectiveness in producing Na-rich stars. The lack
of detection of multiple populations in star clusters younger than 2 Gyr and less massive than 
$\sim$ 10$^5$ $M_{\odot}$ \citep[see figure 32 in ][]{cs2020} could be a sign of 
that  loss of Na enhancement. Indeed, cluster mass would not seem to drive multiple 
population as very old globular clusters less massive than 10$^5$ $M_{\odot}$ harbor multiple
populations. 

Type II globular clusters are characterized by having additional sequences in the chromosome
map in comparison with Type I globular clusters. These additional sequences consist of
red giant branch stars which can have Na abundances similar or higher than the so-called
second generation stars. For this reason, the total Na enhancement in these globular clusters
could be a little higher
than the respective (2G-1G) value. Type II globular clusters are also enhanced in metallicity 
and,  in some cases, $s$-process elements. These characteristics refer to a relatively faster 
nucleosynthesis process than that that could take place in Type I globular clusters, which 
in turn match the features of accreted globular clusters found in our diagnostic diagrams. As can be seen, 
all Type II globular clusters in our sample (represented by large open red circles) have Na
enhancements higher than 0.3 (see Figure~\ref{fig:fig1}), and half of them have retrograde 
orbital motions (see Figure~\ref{fig:fig3}).

Na enhancement has been the most frequent observational evidence to assess on the existence 
of multiple populations. We show here that the amplitude of such a Na enrichment could be
linked with the powerful strength deployed in the early Universe that soon after has become in
a more quiescent nucleosynthesis activity. Although multiple populations seem to 
arise from within the star clusters as a results of intra-cluster processes  (e.g., interactions 
between stars), the star clusters at the time of formation of first and second generation stars 
would seem to keep memory of that cosmological vitality. Therefore, host galaxies would play a
role in the existence of the multiple population phenomenon. We note that \citet{miloneetal2020} did not find any
significant difference in the multiple populations between star clusters associated with different 
progenitors \citep[see, also,][]{saracinoetal2020}. However, there has been a number of
numerical and observational works attempting to describe the formation of globular clusters
with multiple populations that are in very good agreement with some aspects of the intra-cluster
Na enrichment scenario suggested in this work \citep[see, e.g.,][]{bekki2006,carrettaetal10,maxwelletal2014,battagliaetal2017,santistevanetal2020}.

\begin{figure*}
\includegraphics[width=\textwidth]{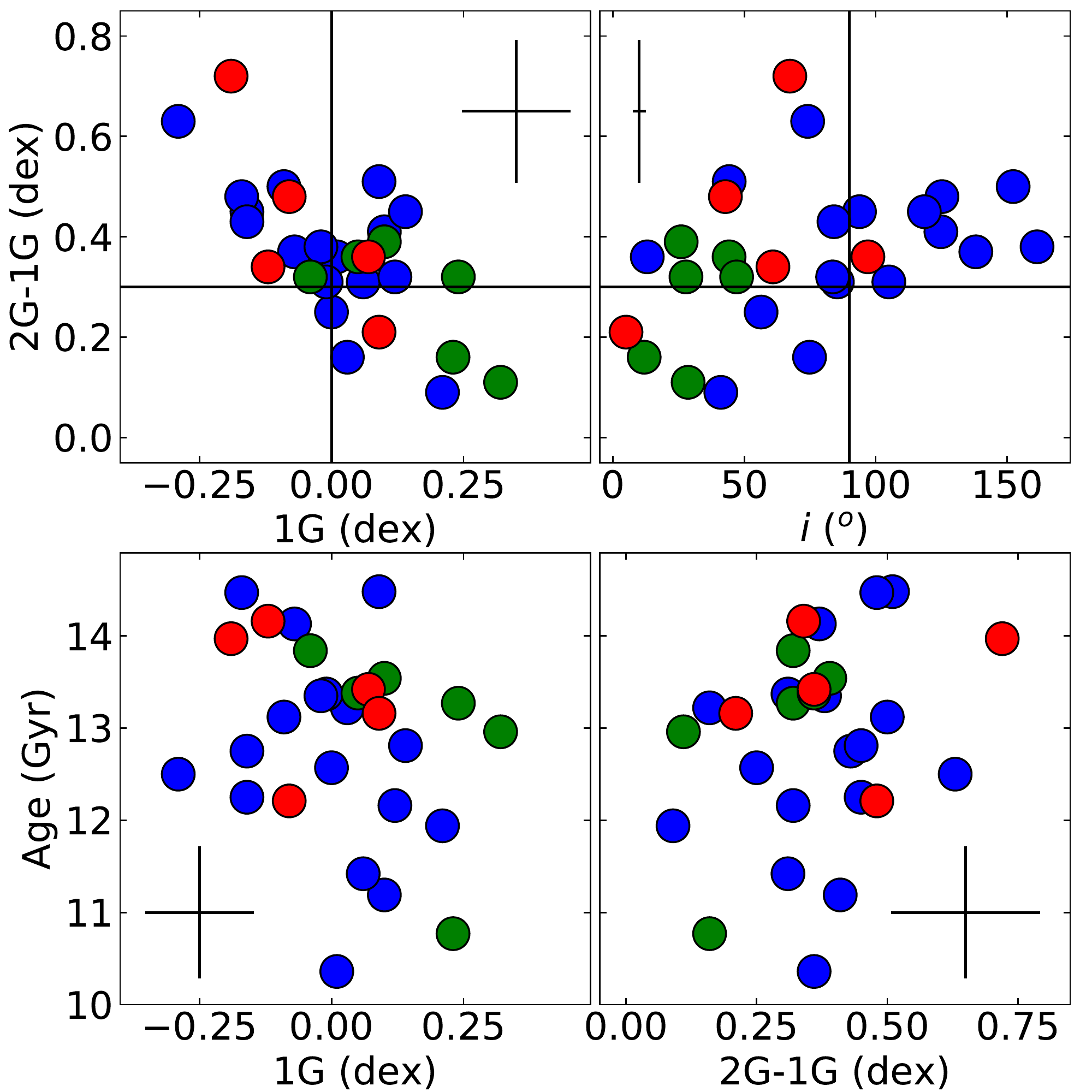}
\caption{Relations between Na abundances of first and second generation stars, the
inclination of the globular clusters' orbits ($i$), and their ages. Globular clusters with an
accreted origin from known large and  small satellites, and formed  in situ are
represented by filled  blue, red, and green circles, respectively.
Typical error bars are included.}
\label{fig:fig5}
\end{figure*}

We distinguish globular clusters with an accreted origin or formed in situ based on 
a combination of their kinematics (prograde versus retrograde orbits) and the Na abundandes
of first and second generation stars. Recently, a fairly substantial literature has dealt with
the classification of Milky Way globular clusters according to the progenitors to which they
could be associated. Table~\ref{tab:tab1} shows such a compilation of possible progenitors. 
The last column lists the status of the globular clusters' origins adopted in this work. 
\citet{piatti2019} discussed extensively the different classifications of Table~\ref{tab:tab1} 
showing that there is some overlap in the list of globular clusters associated to each progenitor.
Another aspect worth of mentioning is that  among the globular clusters associated to a particular 
progenitor, we find those with prograde and retrograde orbits, which means that either the 
selection of globular clusters associated to accreted dwarf galaxies based only on their 
angular momentum, their energies, or on age-metallicity relationships, is not sufficient 
selection criteria. These astrophysical properties, in addition to other properties, would seem 
to be needed. The ratio of accreted to in situ globular clusters is also different in those
studies, so that it is still an open question whether the accreted globular clusters have been
shaped by minor mergers or by one major merger event. Despise the above constraints, it is
still useful to explore whether the results found in this work can globally tracked considering the
adopted progenitors of Table~\ref{tab:tab1}. Figure~\ref{fig:fig5} depicts the relationships
between 1G and 2G-1G with $i$ and the globular clusters' ages as in Figs.~\ref{fig:fig1}-\ref{fig:fig4}.
In this opportunity we painted with blue, red, and green filled circles globular clusters
associated to an accreted large and small satellite and formed in situ, respectively. As can be
seen, there is a broad correspondence that supports the present outcomes, in the sense that
larger Na enhancement are seen in globular clusters associated to accreted satellites.

\begin{figure*}
\includegraphics[width=\textwidth]{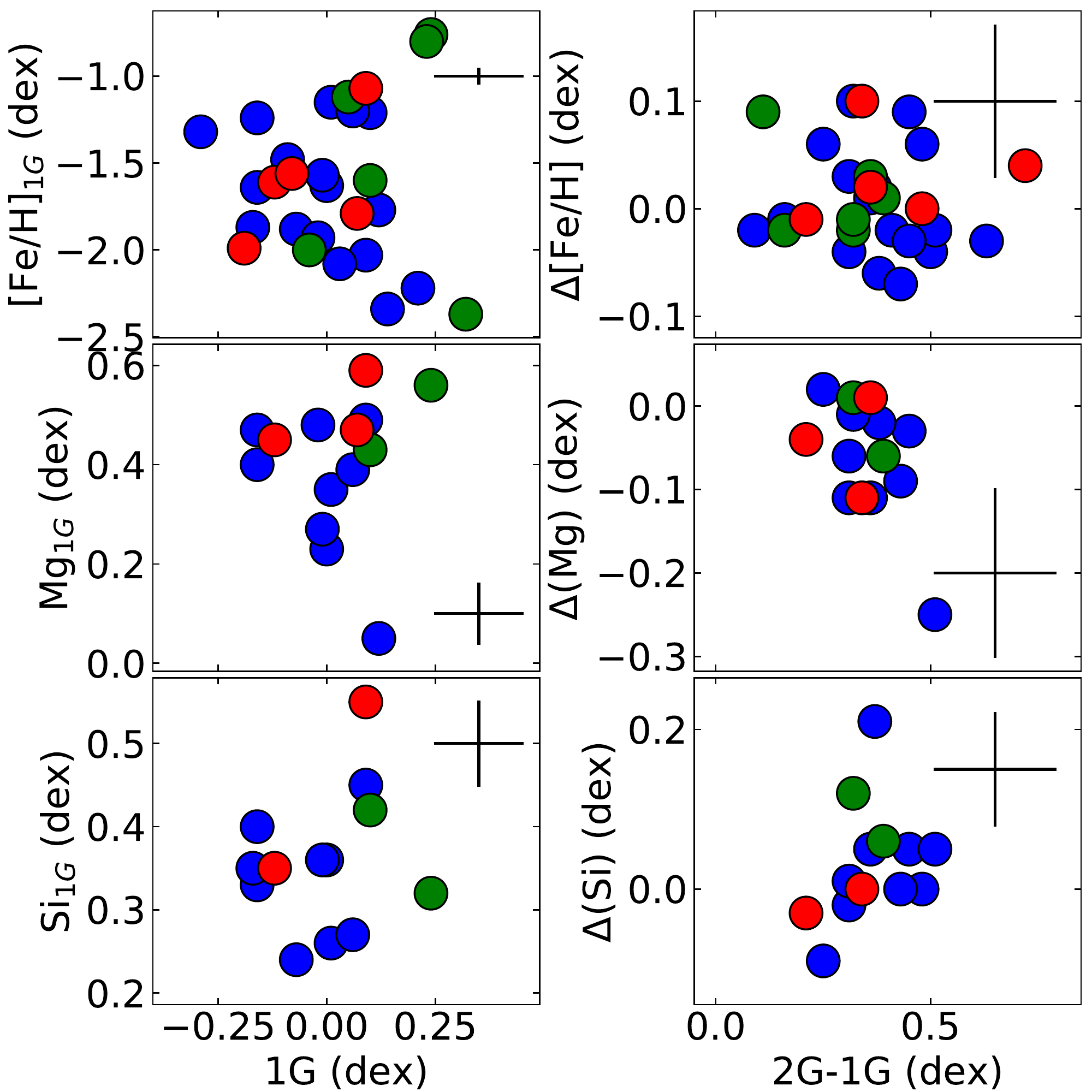}
\caption{Relations between Fe, Mg, Si and Na abundances of first and second generation stars.  
Symbols are as in Fig.~\ref{fig:fig5}. Typical error bars are included. We point out
as a caveat the small number of in-situ globular clusters in the studied sample. Likewise, 
we refer to \citet{ns2010} where the reader might find some support for the idea that 
[Na/Fe] is related to an accretion origin.}
\label{fig:fig6}
\end{figure*}

\citet{tolstoyetal2009} showed that the [Na/Fe] ratio varies as a function of [Fe/H] even
in Milky Way field stars, and can be significantly sub-solar at moderately low metallicities. 
\citet{carrettaetal2009a} showed that the minimum [Na/Fe] in globular clusters follows this 
trend quite well. With the updated compilation of [Fe/H] abundances for first and second
generations stars by \citet{marinoetal2019} we built Fig.~\ref{fig:fig6} (top panels), which
shows that such a correlation is confirms at some extent for globular clusters with a large
satellite progenitor. Globular clusters formed in situ would not seem to exhibit a similar
behavior. Likewise, the difference of [Fe/H] values between first and second generation stars
results independent of the Na enhancement (top-right panel of Fig.~\ref{fig:fig6}).
Variations in [Na/Fe] in field stars are usually correlated with differences in the abundances 
of other elements, notably the alpha-elements (Mg, Si, etc) \citep[see, e.g.,][]{hortaetal2020}. We
here probe such a trend with the Mg and Si abundances available for a subsample of the
studied globular clusters \citep{marinoetal2019}. As can be seen in Fig.~\ref{fig:fig6}
(middle and bottom panels),  it would seem that this is not the case for the present
Milky Way globular cluster sample.

\begin{figure}
\includegraphics[width=\columnwidth]{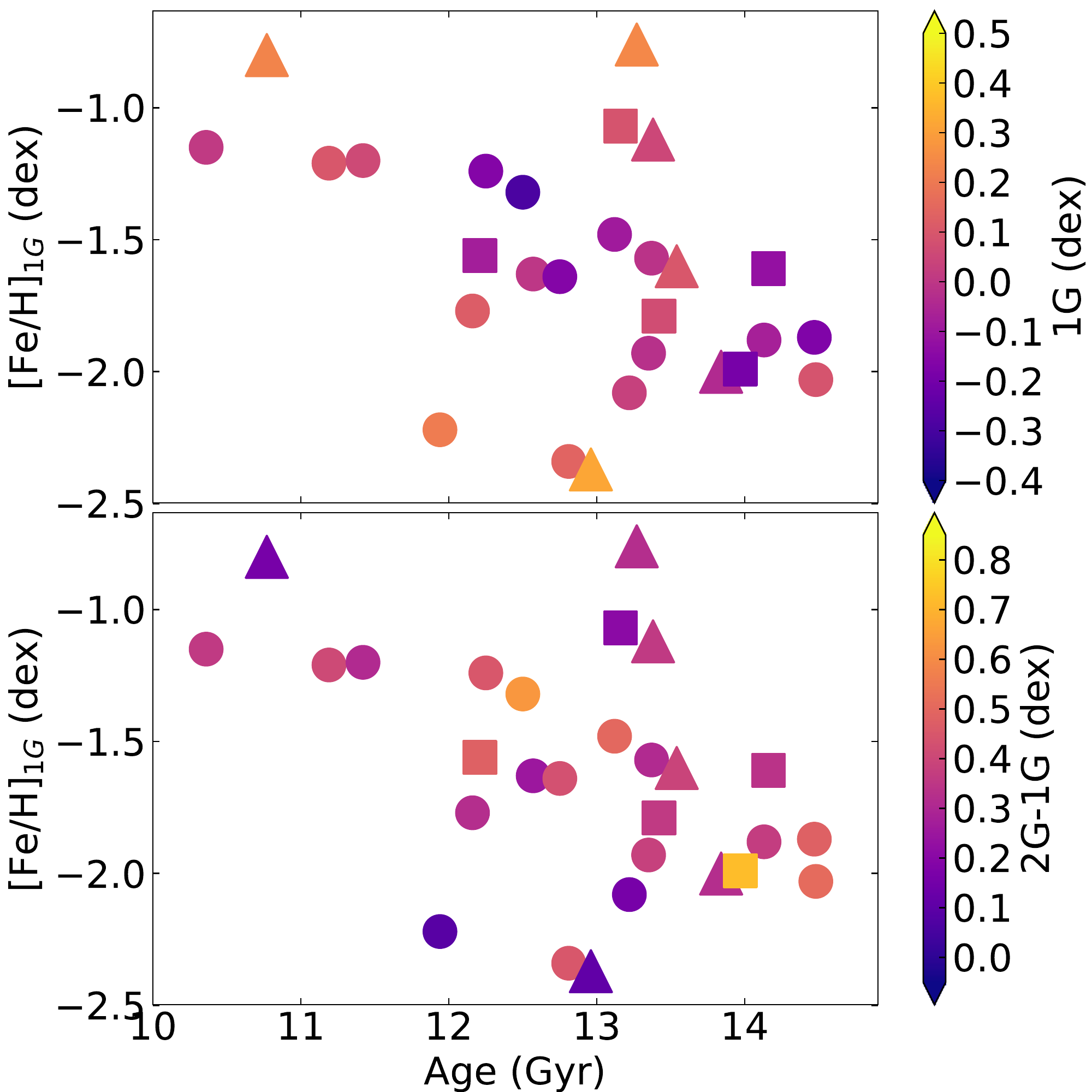}
\caption{Age-metallicity relationship for the studied globular cluster sample. Filled
circles, squares, and triangles represent globular clusters associated to large and small
accreted satellites and formed in situ, respectively. Color-coded symbols represent
1G (top panel) and 2G-1G (bottom panel) values.}
\label{fig:fig7}
\end{figure}

For completeness purposes, we examined the age-metallicity relationship of the studied
globular clusters. We included the entire globular cluster sample in the same plot, although 
different age-metallicity relationships have been invoked in order to recognize globular
clusters associated to different progenitors 
\citep{kruijssenetal2019,massarietal2019,forbes2020}. The resulting age-metallicity relationship
is shown in Fig~\ref{fig:fig7}, where the progenitor status of Table~\ref{tab:tab1} was
considered. Globular clusters associated to large and small accreted satellites and formed
in situ are represented by filled circles, squares and triangles, respectively. Figure~\ref{fig:fig7}
shows a combination of the outcomes illustrated in Figs.~\ref{fig:fig1}, \ref{fig:fig2} and
\ref{fig:fig6} (top panels). It reveals that the most Na-poor globular clusters (1G < 0.0 dex, 
see Fig.~\ref{fig:fig2}) are older than $\sim$ 12 Gyr, and most of them have assigned an 
accreted origin (see Fig.~\ref{fig:fig5}, bottom-left panel). We note that most of the globular 
clusters formed  in situ, regardless their ages, are among those with Na-rich values 
(1G > 0.0 dex, see Fig.~\ref{fig:fig5}), a feature also seen in younger globular clusters 
($\la$ 12 Gyr ) with an accreted origin. Globular clusters formed in situ span the whole age range
and do not follow a tight age-metallicity relationship (see also Figs.~\ref{fig:fig5} and \ref{fig:fig6}
top panels). The bottom panel of Fig.~\ref{fig:fig7}  shows that globular clusters formed in situ
show in general  low Na enhancements.

\begin{table*}
\caption{Origin of Milky Way globular clusters}
\label{tab:tab1}
\begin{tabular}{@{}llcc}\hline\hline
Star cluster &  Progenitor  & Adopted & orbit's inclination ($\degr$) \\\hline
NGC\,104 & main disk (4) & in situ & 27.8\\
NGC\,288 & Gaia-Enceladus (1,6) & large satellite & 124.8\\
NGC\,362 & Gaia-Enceladus (1,6), Kraken (2), Gaia-Sausage (3) &	large satellite & 85.4\\
NGC\,1851 & Gaia-Enceladus (1,6), Canis Major (2); Gaia-Sausage (3) & large satellite & 93.8\\
NGC2808 & Canis Major (2), Gaia-Sausage (3), Gaia-Enceladus (4,6) &large satellite & 13.1 \\
NGC\,3201 & Kraken (2), Sequoia (5,6), Gaia-Enceladus/Sequoia (4) &large satellite & 152.3 \\
NGC\,4590 & Canis Major (2), Helmi streams (4,6) & large satellite & 41.0	\\
NGC\,4833 & Gaia-Enceladus (1) &	large satellite & 44.2\\
NGC\,5024 & Helmi streams (4,6) &large satellite & 74.8	\\
NGC\,5139 & Gaia-Enceladus (1), Kraken (2), Sequoia (5,6) &	large satellite & 138.1\\
NGC\,5272 & Kraken (2), Helmi streams (4,6) &large satellite & 56.4	\\
NGC\,5286 & Gaia-Enceladus (1,6), Canis Major (2), Gaia-Sausage (3) &	large satellite & 125.2\\
NGC\,5904 & Kraken (2), Helmi streams (4,6) / Gaia-Enceladus (4) &large satellite & 74.1	\\
NGC\,5986 & low-energy (4), Koala (6) &	small satellite & 60.9\\
NGC\,6093 & low-energy (4), Koala (6) & small satellite & 97.0\\
NGC\,6121 & los-energy (4), Kraken (2) &	small satellite & 5.0 \\
NGC\,6205 & Gaia-Enceladus (1,6), Canis Major (2) &large satellite & 105.0	\\
NGC\,6254 &  low-energy (4), Koala (6) &	small satellite & 42.8\\
NGC\,6362 & main disk (4) &	in situ & 44.2\\
NGC\,6397 & main disk (4) &	in situ & 47.1\\
NGC\,6535 & Sequoia (5,6) / low-energy/Sequoia (4) &	large satellite & 161.4\\
NGC\,6715 & Saggitarius (2,6) &large satellite	& 83.6\\
NGC\,6752 & Kraken (2), main disk (4) &	in situ & 26.0\\
NGC\,6809 & low-energy (4) &	small satellite & 67.3\\
NGC\,6838 & main disk (4) &in situ & 11.9	\\
NGC\,7078 & Canis Major (2), main disk (4) & in situ	& 28.6\\
NGC\,7089 & Gaia-Enceladus (1,6), Kraken (2), Gaia-Sausage (3) &	 large satellite & 84.1\\
NGC\,7099 & Gaia-Enceladus (1,6) & large satellite & 118.5 \\\hline
\end{tabular}

Ref. : (1) \citet{helmietal2018}; (2) \citet{kruijssenetal2019}; (3) \citet{myeongetal2018}; 
(4) \citet{massarietal2019}; (5) \citet{myeongetal2019}; (6) \citet{forbes2020}.

\end{table*}

\begin{acknowledgements}
I thank the referee for the thorough reading of the manuscript and
timely suggestions to improve it. 
\end{acknowledgements}



\begin{appendix} 

\section{Na abundances of Milky Way globular clusters}

\begin{figure}
\includegraphics[width=\columnwidth]{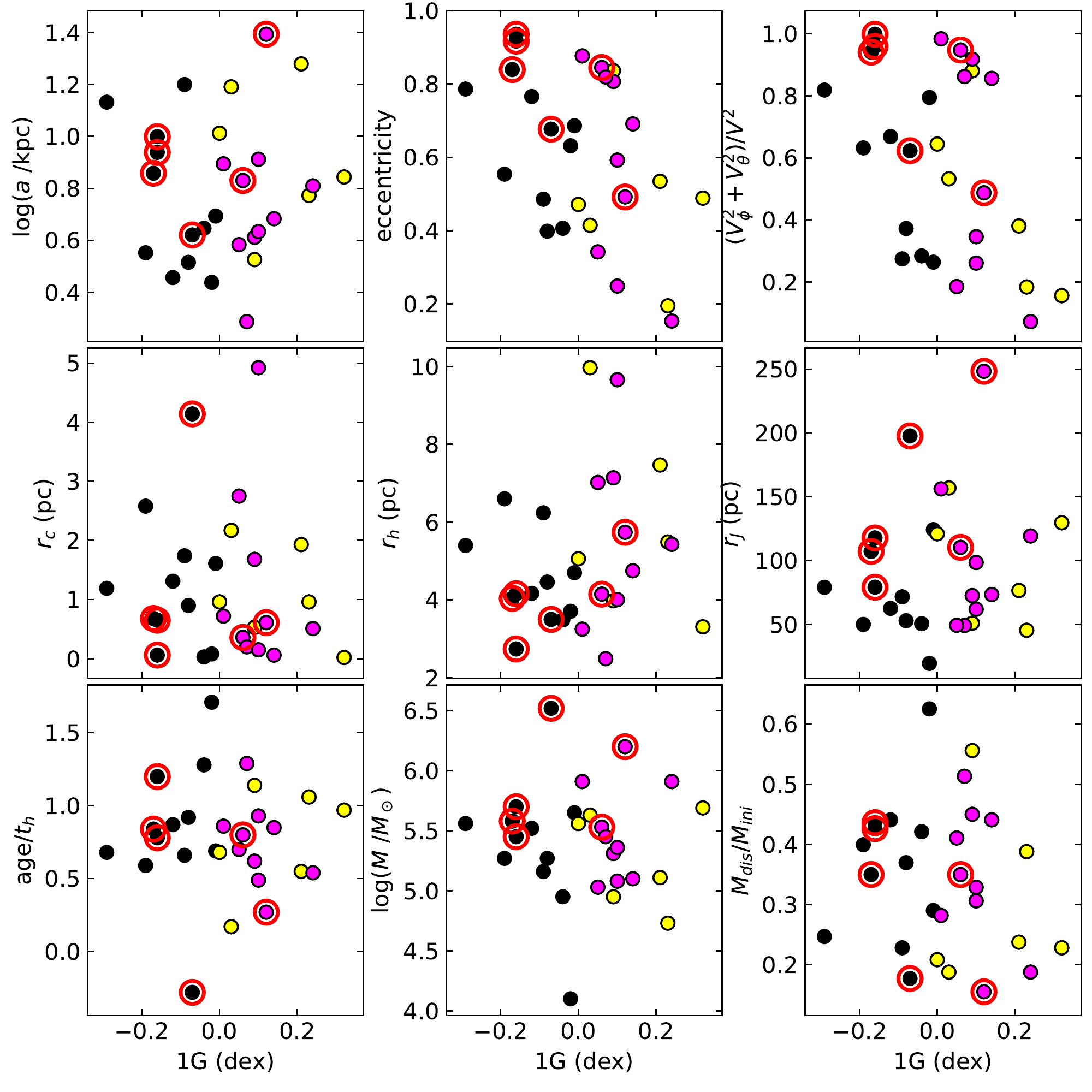}
\caption{Na abundances of first generation stars as a function of different
globular clusters' parameters, namely: semi-major axis ($a$) of the globular
cluster orbits, eccentricity of the globular cluster orbits, space velocity components
$V_\phi$ and $V_\theta$, $V^2$= $V_\phi^2$+$V_\theta^2$+$V_r^2$, core radius
($r_c$), half-mass radius ($r_h$), Jacobi radius ($r_J$), age to half-mass relaxation
time ratio (age/$t_h$), globular cluster mass, and ratio of the mass lost by tidal disruption
to the total globular cluster mass ($M_{dis}/M_{ini}$). }
\label{fig:figa1}
\end{figure}

\begin{figure}
\includegraphics[width=\columnwidth]{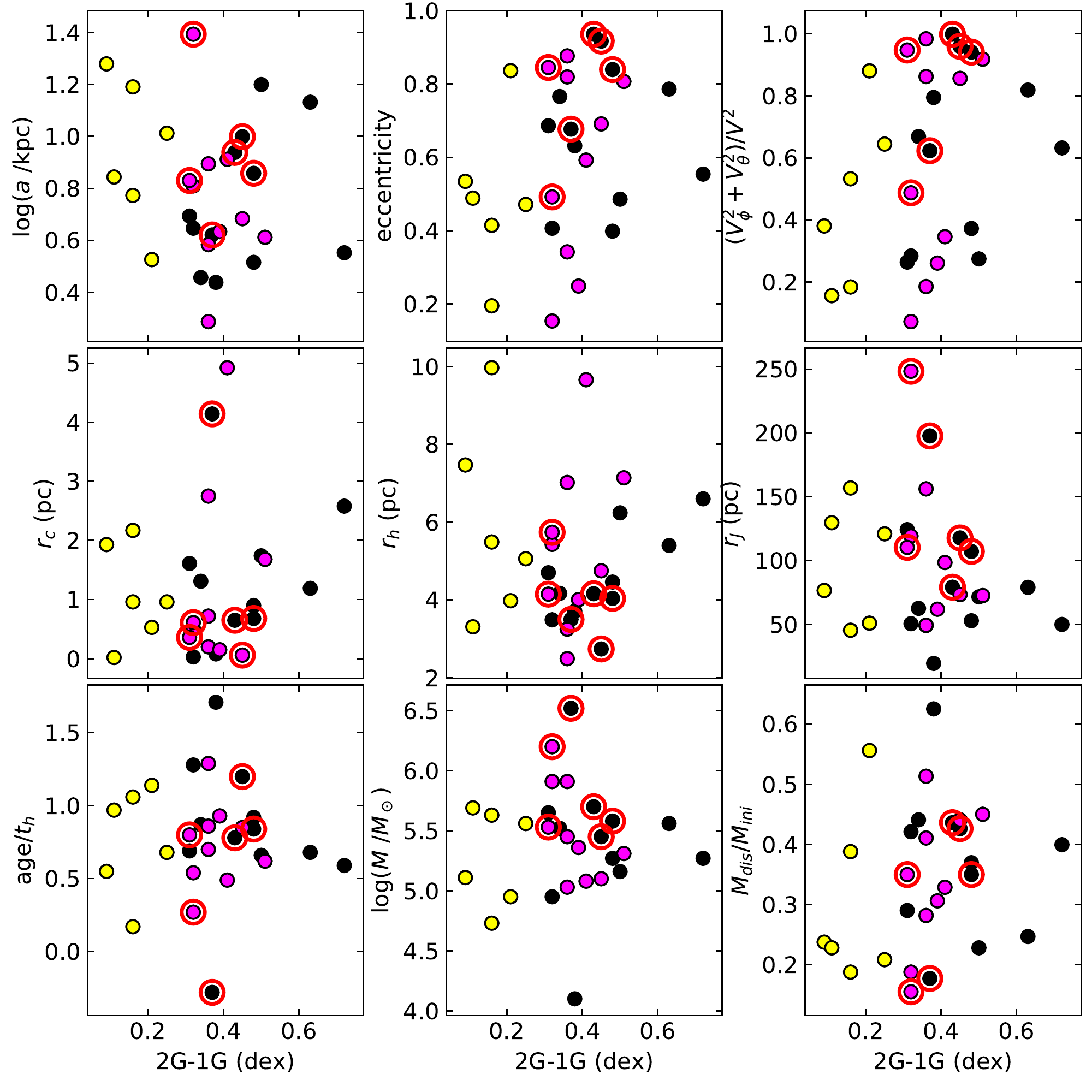}
\caption{Same as Fig.~\ref{fig:figa1} for the difference of Na abundances
between first and second generation stars.}
\label{fig:figa2}
\end{figure}

In this section we present the relationships of 1G and 2G-1G with
different astrophysical parameters of the studied Milky Way globular
clusters (see Section 2). Symbols are as in Fig.~\ref{fig:fig1}. As can be
seen in Figs.~\ref{fig:figa1}-\ref{fig:figa2}, there is not a clear dependence 
of 1G and 2G-1G with them, except
for the inclination of the globular clusters' orbits, which we used in Sect. 3.

\end{appendix}

\end{document}